\documentclass[12pt,a4paper,oneside]{article}
\usepackage[hypertex]{hyperref}
\usepackage{graphicx,amssymb,amsmath,slashed}
\usepackage[margin=2cm]{geometry}
\begin{document}

\newcommand{\be}{\begin{equation}}
\newcommand{\ee}{\end{equation}}
\newcommand{\bea}{\begin{eqnarray}}
\newcommand{\eea}{\end{eqnarray}}
\newcommand{\beq}{\begin{equation}}
\newcommand{\eeq}{\end{equation}}
\newcommand{\beqa}{\begin{eqnarray}}
\newcommand{\eeqa}{\end{eqnarray}}
\newcommand{\no}{\nonumber}

\newcommand{\cL}{\mathcal{L}}
\newcommand{\cO}{\mathcal{O}}
\newcommand{\cA}{\mathcal{A}}
\newcommand{\cB}{\mathcal{B}}

\newcommand{\eg} {{\it e.g.}}
\newcommand{\ie}  {{\it i.e.}}
\newcommand{\et} {{\it et. al}}

\newcommand{\mtt}{M_{t \bar t}}
\newcommand{\ah}{A_{450}^{t\bar t}}

\title{\fontsize{16}{20}\selectfont\textbf{Implications of the CDF $t\bar t$ Forward-Backward Asymmetry for Hard Top
Physics}}
\author{\fontsize{12}{16}\selectfont C\'edric Delaunay, Oram Gedalia, Yonit Hochberg, Gilad
Perez and Yotam Soreq \vspace{6pt}\\
\fontsize{11}{16}\selectfont\textit{Department of Particle
Physics and Astrophysics, Weizmann Institute of Science,}\\
\fontsize{11}{16}\selectfont\textit{Rehovot 76100, Israel}}
\date{}
\maketitle

\begin{abstract}
The CDF collaboration has recently reported a large deviation
from the standard model of the $t \bar t$ forward-backward
asymmetry in the high invariant mass region. We interpret this
measurement as coming from new physics at a heavy scale
$\Lambda\,$, and perform a model-independent analysis up to
$\cO(1/\Lambda^4)\,$. A simple formalism to test and constrain
models of new physics is provided. We find that a large
asymmetry cannot be accommodated by heavy new physics that does
not interfere with the standard model. We show that a smoking
gun test for the heavy new physics hypothesis is a significant
deviation from the standard model prediction for the $t \bar t$
differential cross section at large invariant mass. At
$M_{t\bar t}>1$~TeV the cross section is predicted to be at
least twice that of the SM at the Tevatron, and for $M_{t\bar
t}>1.5$~TeV at least three times larger than the SM at the LHC.
\end{abstract}

\section{Introduction}
The top quark is the most massive point-like particle known to
exist. As a consequence, within the Standard Model (SM), the
top is largely responsible for the hierarchy problem.
Furthermore, in most natural models it is linked to electroweak
symmetry breaking. Therefore, there is strong motivation to
search for new physics effects associated with top physics.

The CDF collaboration has recently announced several intriguing
new measurements that exhibit large deviations from the
corresponding SM predictions. Evidence for an anomalous
forward-backward $t\bar t$ production asymmetry was observed
for large invariant mass of the $t\bar t$
system~\cite{Aaltonen:2011kc}:
\beq\label{eq:atthexp}
A^{t\bar t}_{450}\equiv A^{t\bar t}(M_{t\bar t}\geq450\ {\rm
GeV})=+0.475\pm0.114\,,
\eeq
to be compared with the SM prediction~\cite{Bowen:2005ap,
Antunano:2007da, Almeida:2008ug}, $A^{t\bar
t}_{450}=+0.09\pm0.01$. Previous D0 and CDF measurements of the
inclusive $t \bar t$ asymmetry~\cite{:2007qb, Aaltonen:2008hc}
also show deviation from the SM prediction. Another
recent CDF analysis in the dilepton channel~\cite{cdfdilepton}
supports this deviation, and furthermore finds a rising $\mtt$
dependence for the forward-backward asymmetry.

Additionally, the CDF collaboration has recently made progress
in studying the mass distribution of highly boosted jets
($p_T>400$~GeV for the leading jet)~\cite{cdfboostedqcd}, and
found a hint for an excess of events in the high mass
region~\cite{cdfboostedtops}.

The above measurements suggest that new physics affecting the
top sector is present. Our approach in this work is the
following. We interpret the measurement of $A^{t\bar t}_{450}$
in terms of new physics, checking the consistency of such a
scenario with other measurements that do not show any
significant deviation from the SM predictions. We further
discuss the effects of such new physics on ultra-massive
boosted jets at the Tevatron. We then make predictions for the
invariant mass distributions of top pairs soon to be measured
at the Tevatron and LHC.

Several works have interpreted the recent CDF measurement of
$A^{t\bar t}_{450}$ within specific models of new
physics~\cite{Cheung:2011qa,Delaunay:2011vv, Cao:2011ew,
Bai:2011ed,Shelton:2011hq,Berger:2011ua,Gresham:2011dg,
Barger:2011ih, Bhattacherjee:2011nr,Grinstein:2011yv,
Patel:2011eh,Isidori:2011dp, Zerwekh:2011wf, Barreto:2011au}.
Similarly, model-independent analyses were
performed~\cite{Jung:2009pi, Degrande:2010kt,Jung:2010yn} and
new physics models were invoked~\cite{Ferrario:2008wm,
Ferrario:2009bz,Jung:2009jz, Cheung:2009ch, Frampton:2009rk,
Shu:2009xf,Arhrib:2009hu, Dorsner:2009mq, Cao:2009uz,
Barger:2010mw,Cao:2010zb, Xiao:2010hm,Chivukula:2010fk,
Chen:2010hm, Alvarez:2010js} to explain earlier D0 and CDF
measurements of the inclusive asymmetry~\cite{:2007qb,
Aaltonen:2008hc}.

We focus on the class of models in which the scale of new
physics is well above the scale $\mtt$ relevant to the CDF
measurements. The effects of such new physics can then be
described from a low energy model-independent perspective,
using the language of effective field theory.
Ref.~\cite{ourfirst} performed a similar analysis, further
assuming that the dominant contribution to the forward-backward
asymmetry comes from interference between the new physics and
SM contributions to top pair production. Denoting the scale of
new physics by $\Lambda$, Ref.~\cite{ourfirst} found that in
the presence of an axial octet operator producing a pair of
tops from a pair of up quarks at ${\cal O}(1/\Lambda^2)\,$, the
observed $t\bar t$ forward-backward asymmetry can be accounted
for. Here we relax the assumption of interference and consider
all operators contributing to $t\bar t$ production up to order
$1/\Lambda^4$. We provide a simple formalism that enables one
to easily obtain constraints and predictions for models
consistent with our framework. We derive model-independent
predictions regarding near future measurements that will
sharply test our general underlying assumptions.

The paper is organized as follows. In Section~\ref{sec:data} we
review the data relevant to our study. Section~\ref{sec:op}
defines the set of operators in our effective Lagrangian.
Section~\ref{sec:rel} relates the operators to the observables.
Our results are presented in Section~\ref{sec:afb}. In
Section~\ref{sec:pred} we discuss predictions for hard top
physics at the Tevatron and LHC. We conclude in
Section~\ref{sec:conc}.

\section{Relevant Data}\label{sec:data}

In this work we analyze the effect of heavy new physics on the
forward-backward asymmetry at large $\mtt$. Roughly, we aim to
account for a new physics contribution of
\be \label{afb_range}
A^{t \bar t}_{450}=+0.40 \pm 0.11\, ,
\ee
assuming that the rest of the asymmetry in
Eq.~\eqref{eq:atthexp} comes from the SM.

Other top-related measurements do not show significant
deviations from the SM predictions. Consequently, they provide
constraints on the parameter space of the effective
Lagrangian. The first such observable is the $t \bar t$
differential cross section, which we choose to represent by the
following large $\mtt$ bin~\cite{Aaltonen:2009iz}
\be \label{sigmah}
\sigma_{700}\equiv \sigma^{t\bar t}(700\ {\rm GeV}<M_{t\bar
t}<800\ {\rm GeV})=80\pm37\ {\rm fb}\,,
\ee
as in~\cite{ourfirst}. This is consistent with the SM
prediction~\cite{Almeida:2008ug,Ahrens:2010zv},
$\sigma_{700}=80\pm8$ fb. In~\cite{ourfirst}, the inclusive $t
\bar t$ cross section was also used as a constraint. However,
the theoretical estimation of the cross section originating
from threshold effects is still under investigation
(compare~\cite{Cacciari:2008zb,
Kidonakis:2010dk,Langenfeld:2009wd} with~\cite{Ahrens:2010zv}).
Furthermore, the dynamics of our heavy new physics naturally
affects the measurement at large invariant masses more
significantly. Thus in our study we do not use the inclusive
$t\bar t$ cross section to constrain our parameter space. (Note
though that our results below are within the combined
theoretical and experimental uncertainties for the inclusive
observables.) The same argument leads us to refrain from
considering $A^{t \bar t}$ in the low invariant mass region, as
well as the inclusive asymmetry. We therefore use a
theoretically-cleaner observable, also relevant for $A^{t \bar
t}_{450}\,$, related to the cross section above 450~GeV,
$\sigma_{450}\,$. Note that there seems to be some discrepancy
in~\cite{Aaltonen:2009iz} between the measurement in this range
and the SM prediction. However, in a more recent measurement
reported in~\cite{Aaltonen:2011kc} (but not translated to the
partonic level) this discrepancy is not present. We thus assume
that the SM prediction agrees with the measured value of the
cross section above 450~GeV. For concreteness, we use the
relative uncertainty reported in~\cite{Aaltonen:2011kc} for the
450-500~GeV $\mtt$ bin (see below), which dominates the
uncertainty in the $\mtt>450$~GeV range.

In order to minimize the impact of next to leading order (NLO)
corrections to the new physics (NP) contributions, we normalize
the latter to the SM one. We assume that the $K$-factors are
universal, so that the NP/SM ratios at leading order and next
to leading order are the same. This assumption is reasonable for the effective operators generated in the SM after the highly virtual gluon is integrated out~\cite{ourfirst}. Additional NLO corrections are formally down by ${\cal O}(\alpha_s/\pi)$ and are henceforth neglected. In practice, $\log$-enhancements of these contributions might be present which could modify the analysis. A full NLO computation of the effective theory is yet to be done, and is beyond the scope of this work. However, as a self consistency check of our analysis, we find below that the SM-like operators indeed account for the dominant part of the forward-backward asymmetry (see Fig.~\ref{fig:cacv}), supporting the above assumption.

Combining in quadrature the
experimental and theoretical uncertainties, we represent
Eq.~\eqref{sigmah} and the uncertainty on the cross section
above 450~GeV as follows:
\be \label{ns}
\left|N_{700}\right|\equiv \left| \sigma^{\rm
NP}_{700}/\sigma^{\rm SM}_{700} \right| \lesssim0.5\,, \qquad
\left|N_{450}\right|\equiv \left| \sigma^{\rm
NP}_{450}/\sigma^{\rm SM}_{450} \right| \lesssim0.2\,.
\ee

It is also intriguing to explore the implications of the new
physics in the context of the CDF boosted jets
study~\cite{cdfboostedqcd,cdfboostedtops}. The cross section
for ultra-massive boosted jets (not coming from QCD events) can
be estimated as follows~\cite{ourfirst}
\beq\label{eq:defsigb}
\sigma_b \sim \left[21- \left(8.7\pm3.1\right) R_{\rm
mass}^{-1}\right] {\,\rm fb},
\eeq
where $\sigma_b$ is the cross section of hadronically-decaying
$t \bar t$ with a $p_T$ cut of 400~GeV on the leading jet and
$R_{\rm mass}$ is a parameter that determines the QCD
background, as defined in~\cite{ourfirst,gluino}. (An
assumption of naive factorization of the jet mass distribution
yields $R_{\rm mass}=1\,$, while matched Monte-Carlo
simulations give $R_{\rm mass}=0.87$~\cite{ourfirst,gluino}
with an excellent agreement on this value between the different
generators.) The SM prediction for the top contribution is
$\sigma^{\rm SM}_b=2.0\pm 0.2$\,fb~\cite{Kidonakis:2003qe}. We
interpret the excess as top pairs, generated by the new physics
source. The magnitude of this effect is then~\cite{ourfirst}
\be\label{eq:Nbexp}
N_b \equiv \sigma_b^{\rm NP}/\sigma_b^{\rm SM}=5 \pm 2 \,,
\ee
where $\sigma_b^{\rm NP}$ is the new physics contribution to
the boosted cross section, assuming $R_{\rm
mass}=0.87$.

\section{The Operator Basis} \label{sec:op}

As stated above, the basic assumption that we employ is that
the new physics is characterized by a scale $\Lambda$ that is
larger than the invariant mass of the top pair $\mtt\,$ in the
measurements which we consider. The natural approach is then to
use a set of effective operators to describe the new physics.
These operators must lead from an initial $u\bar u$ state to a
final $t\bar t$ state, and as such appear at dimension six and
higher. (The contribution of $d\bar d\to t\bar t$ at the
Tevatron is at most $15\%$ that of $u\bar u\to t\bar t$ for
$M_{t\bar t}$ above 450~GeV, as relevant for the observables
that we consider.)

At $\cO(1/\Lambda^2)$, there are only two four-quark operators
that interfere with the SM:
\be \label{dimsixinterfere}
\cO^8_V = \left(\bar{u}\gamma_\mu T^a u\right)
\left(\bar{t}\gamma^\mu T^a t\right)\,, \qquad \cO^8_A =
\left(\bar{u}\gamma_\mu\gamma^5T^a u\right)
\left(\bar{t}\gamma^\mu\gamma^5 T^a t\right)\,,
\ee
where the superscript 8 denotes an octet color structure.
Allowing for contributions that do not interfere with the SM,
there are two more vector octet operators at this order:
\be \label{dimsixvector}
\cO^8_{AV} = \left(\bar{u}\gamma_\mu\gamma^5 T^a u\right)
\left(\bar{t}\gamma^\mu T^a t\right) \,, \qquad \cO ^8_{VA} =
\left(\bar{u}\gamma_\mu T^a u\right)
\left(\bar{t}\gamma^\mu\gamma^5 T^a t\right)\,.
\ee
There are four additional orthogonal combinations of color
contractions, given by:
\be \label{dimsixsing}
\begin{split}
\cO^1_V &= \left(\bar{u}\gamma_\mu u\right)
\left(\bar{t}\gamma^\mu t\right) \,, \qquad \cO^1_A =
\left(\bar{u}\gamma_\mu\gamma^5 u\right)
\left(\bar{t}\gamma^\mu\gamma^5 t\right)\,, \\ \cO^1_{AV}&=
\left(\bar{u}\gamma_\mu\gamma^5 u\right)
\left(\bar{t}\gamma^\mu  t\right) \,, \qquad \cO^1_{VA} =
\left(\bar{u}\gamma_\mu u\right)
\left(\bar{t}\gamma^\mu\gamma^5 t\right)\,.
\end{split}
\ee
The list of dimension six operators is concluded with eight
scalar and two tensor operators:
\be\label{eq:op6ST}
\begin{split}
\cO^{1,8}_S&= \left(\bar{u}\,T_{1,8}u\right)
\left(\bar{t}\,T_{1,8} t\right)\,, \qquad \cO^{1,8}_P =
\left(\bar{u}\,T_{1,8}\gamma^5u\right)
\left(\bar{t}\,T_{1,8}\gamma^5t\right)\,, \\ \cO^{1,8}_{SP}&=
i\left(\bar{u}\,T_{1,8}u\right) \left(\bar{t}\,T_{1,8}\gamma^5
t\right)\,, \qquad \cO^{1,8}_{PS} =
i\left(\bar{u}\,T_{1,8}\gamma^5u\right)
\left(\bar{t}\,T_{1,8}t\right)\,, \\ \cO^{1,8}_T &=
\left(\bar{u}\,T_{1,8}\sigma^{\mu\nu}u\right)
\left(\bar{t}\,T_{1,8}\sigma_{\mu\nu}t\right)\,,
\end{split}
\ee
with $T_1\equiv 1$ and $T_8\equiv T^a\,$.

The above dimension six operators contribute to top pair
production at ${\cal O}(1/\Lambda^4)$ as well, via the square
of their amplitudes. Another type of contribution at ${\cal
O}(1/\Lambda^4)$ comes from chirality-conserving dimension
eight operators that interfere with the SM. These can be
constructed by applying two covariant derivatives in various
ways to the operators in Eq.~\eqref{dimsixinterfere}. However,
naive dimensional analysis shows that their value is given by
$c^2/(16 \pi^2)\,$, where $c$ is the coefficient of any
dimension six operator. This condition generically holds in
case of strongly-coupled new physics, {\it i.e.}\ when the NP
scale is roughly 5-10~TeV~\cite{ourfirst}. Moreover, even in
models with a lower scale (\eg\ a $\sim$2~TeV $s$-channel
resonance), producing a large value for $\ah$ typically leads
to a suppression of the dimension eight contributions compared
to the square of the dimension six amplitudes. We thus neglect
these dimension eight contributions in what follows.

Note that in principle there are also dimension six
chromo-magnetic/electric $u$ and $t$ dipole operators that can
be considered. Their effects at $\cO(1/\Lambda^2)$ in the hard
$\mtt$ regime were shown to be negligible in~\cite{ourfirst}.
As they involve chirality flips, their contributions at order
$1/\Lambda^4$ are suppressed by at least $(m_t/\Lambda)$
compared to their $1/\Lambda^2$ effects. There are also chirality-flipping
dimension eight operators which interfere with the SM, and are
again suppressed by the same factor.

To conclude, we describe the hard region of the $t\bar{t}$
physics by the following effective Lagrangian:
\be
\cL_{\rm eff}=\sum_i \frac{c_i}{\Lambda^2}\mathcal{O}_i \,,
\ee
where the $c_i$ are real coefficients and the operators $\cO_i$
are listed in Eqs.~\eqref{dimsixinterfere}-\eqref{eq:op6ST}.
Below for simplicity of notation, $c_i$ will denote
$c_i/\Lambda_{\rm TeV}^2$, where $\Lambda_{\rm TeV} \equiv
\Lambda/{\rm TeV}\,$.

In our analysis we perform all
calculations at leading order and neglect renormalization group
running and mixing. Consequently, we also do not discuss the
contribution from operator mixing to dijet production at the
LHC~\cite{Bai:2011ed}. In principle, the NP can couple to light and heavy quarks with different strengths. Generically, the operators discussed above can be characterized by a mixed coupling of NP to light and heavy quarks, $g_{u\bar u}\times g_{t\bar t}$, whereas the constraints from dijets are sensitive to operators characterized by a strength $g_{u\bar u}^2$. Examining the present bounds from dijet production at the LHC~\cite{Khachatryan:2011as,Aad:2011aj}, a hierarchy of $g_{u\bar u}/g_{t\bar t}\sim 1/5$ is required in order to comply with the data.

\section{Relating Operators to Data} \label{sec:rel}
We now write the contribution of the operators to the various
observables of interest.  We first focus on the vector operators of
Eqs.~\eqref{dimsixinterfere}-\eqref{dimsixsing}.

It is natural within the vector sector to distinguish between
the operators that interfere with the SM and those that do not.
The latter set of operators can be parameterized by:
\be
\begin{split}
w_\pm^2 \equiv \frac{1}{2}&\left\{\left( c^{8}_{VA} \pm
c^{8}_{AV}\right)^2+\frac92 \left[ \left( c^1_V\pm
c^1_A\right)^2 +\left( c^{1}_{VA}\pm
c^{1}_{AV}\right)^2\right]\right\}\,, \\
&\qquad R^2 \equiv w_+^2+w_-^2 \,, \qquad \tan \theta \equiv
w_-/w_+ \,.
\end{split}
\ee
The relevant observables of Sec.~\ref{sec:data} then take the
simple form
\bea
N_X&\simeq&a_X c_V^8 + b_X (c_V^8)^2 +d_X (c_A^8)^2+
e_X R^2\,, \label{NXfinal}\\
A^{t \bar t}_{450}&=&\left(\alpha c_A^8+\beta c_A^8c_V^8
+\frac{\beta}{2} R^2\cos2\theta\right)
\left(1+N_{450}\right)^{-1}\,,\label{AFBfinal}
\eea
where the subscript $X=450,750,b$ and in $N_X$ we neglect a
term which is proportional to $\sin 2\theta$ and suppressed by
$4m_t^2/\mtt^2\,$. The coefficients $(a,b,d,e)_X$ and
$\alpha,\beta$ are pure kinematical factors given by
\bea
(a,b,d,e)_{450}&=&0.35,\,0.043,\,0.023,\,0.033 \,, \\
(a,b,d,e)_{700}&=&0.76,\,0.16,\,0.11,\,0.14 \,, \\
(a,b,d,e)_b&=&1.5,\, 0.57,\, 0.46,\, 0.51 \,,
\\ \alpha,\beta&=&0.17,\,0.043\,,\label{numbers}
\eea
where we use the MSTW parton distribution
functions~\cite{Martin:2009iq} at leading order in this
calculation. The physical interpretation of $R$ is very
clear~--- it parameterizes the overall size of the operators
which do not interfere with the SM. The angle $\theta$ controls
how much these operators project on the asymmetry. For a given
$R$, the asymmetry is maximized for $\cos2\theta=1$, justifying
the omission of the $\sin 2\theta$ term above.

It is useful to obtain relations between the various
observables, allowing for a simple estimation of the new
physics contributions in terms of the constraints. One such
relation is between the boosted tops enhancement factor and the
two cross section constraints, given by
\be \label{nrelation}
N_b=-0.12\, c^8_V+5.6\, N_{700}-7.5\, N_{450} \,,
\ee
such that
\be \label{eq:cvcon}
\left| c^8_V+10 \right|< 10 \sqrt{1+N_{450}-0.21\,N_{700}}
\quad {\rm and} \quad \left| c^8_V+21 \right|>15.6
\sqrt{1.8+N_{450}-0.24\,N_{700}} \,.
\ee
The constraints in Eq.~\eqref{eq:cvcon} define the range in
which $R \geq0$ and $c_A^8$ is real.

Another useful relation is between $A_{450}^{t \bar t}$ and the
constraints $N_{450}$ and $N_{700}$. To obtain this, we can
substitute $c_A^8$ for $N_{700}$ using Eq.~\eqref{NXfinal},
yielding
\be
(1+N_{450})\times A^{t \bar t}_{450}=
\left(0.51+0.13\,c_V^8\right)\sqrt{N_{700}-\left[0.76\,
c_V^8+0.16\,(c_V^8)^2+0.14\,R^2\right]}+0.022\,R^2\cos2\theta\,.
\ee
This relation is only valid for $c_A^8 \geq 0\,$, and indeed,
as we show below, accounting for $A_{450}^{t \bar t}$ as in
Eq.~\eqref{afb_range} requires $c_A^8
>0\,$.

We would now like to comment on the most general case where the
scalar and tensor operators of Eq.~\eqref{eq:op6ST} are also
present. It is straightforward to show that the effects of the
latter can be captured by our
Eqs.~\eqref{NXfinal}-\eqref{numbers} with the following
redefinitions
\bea
\cos2\theta &\to& F(\theta,\psi,\theta_V,\theta_{ST})\equiv
\cos2\theta\cos^2\theta_V+\sqrt{\frac23}\cos2\psi\sin^2\theta_V\cos^2\theta_{ST}\,,\\
R^2&\to& R^2\equiv w_+^2 + w_-^2 + r^2_{ST} + r^2_P\,,
\eea
with $|F|\leq 1$, and where we defined
\bea
r_{ST}^2&\equiv & y_+^2 + y_-^2\,,\quad y_\pm^2 \equiv \left(c^8_T
\pm \sqrt{\frac38} c_S^8\right)^2+\frac{9}{2}\left(c^1_T \pm \sqrt{\frac38} c_S^1\right)^2\,,\\
r_P^2&\equiv & \frac34\left\{(c_P^8)^2+(c_{SP}^8)^2+(c_{PS}^8)^2
+\frac92 \left[(c_P^1)^2+(c_{SP}^1)^2+(c_{PS}^1)^2\right]\right\}\,,
\eea
and
\be
\sqrt{w_+^2 + w_-^2}\equiv R\cos\theta_V\,,\ r_{ST}\equiv R\sin\theta_V\cos\theta_{ST}\,,\
r_P\equiv R\sin\theta_V\sin\theta_{ST}\,,\ \tan\psi \equiv  {y_+/ y_-}\,.
\ee
The physical interpretation of the additional parameters is as
follows. $r_{ST,P}$ represent the overall size of the
scalar/tensor and pseudo-scalar operators; $R$ being now the
overall size of all of the non-interfering operators. The
angles $\theta_{V,ST}$ determine the distribution of $R$ among
the vector, scalar/tensor and pseudo-scalar operators, while
the angle $\theta$ ($\psi$) parameterizes how much the vector
(scalar/tensor) operators project on the asymmetry. Note that
in the case where vector operators are absent, one has $|F|\leq
\sqrt{2/3}$.

\section{The Forward-Backward Asymmetry} \label{sec:afb}
The stage is now set to study the parameter space that explains
the large forward-backward $t \bar t$ asymmetry $\ah$ while
satisfying the constraints from the differential cross section.
We further comment on the contribution to boosted top pair
production. We consider several interesting limiting cases. The
first is when there is no interference with the SM. Next the
cases where the interference comes from only one operator are
analyzed. Finally, we discuss the general scenario.

\subsection{No Interference: $c^8_A=c^8_V=0$}
In models of heavy new physics which does not interfere with
the SM, a simple relation between $A_{450}^{t \bar t}$ and
$N_{700}$ can be obtained from Eqs.~\eqref{NXfinal}
and~\eqref{AFBfinal}:
\be
\ah=\frac{0.16 \cos2\theta\, N_{700}}{1+0.24\, N_{700}} \,.
\ee
Using the bound on $N_{700}$ from Eq.~\eqref{ns} (which
automatically satisfies the constraint from $N_{450}$), we find
$\ah \lesssim 0.07\,$. From this we learn that the large $t
\bar t$ asymmetry measured by CDF cannot be explained by a
heavy new physics sector which does not interfere with the SM.
This is consistent with the findings
of~\cite{Grinstein:2011yv}, obtained directly from the data.
Additionally, the maximal excess in the high-$p_T$ $t \bar t$
cross section is $N_b \sim2\,$.

\subsection{Vector Interference: $c^8_A=0$ and $c^8_V \neq0$}
It is clear from Eq.~\eqref{AFBfinal} that $\cO^8_V$ by itself
does not contribute directly to $\ah\,$ if $c^8_A=0\,$.
However, taking $c^8_V<0$ can relax the constraints from
$N_{450}$ and $N_{700}$, thus allowing for a larger
contribution to $\ah$ from other operators, \ie~$R \neq 0$.
Substituting $R^2$ for $N_{700}$ and $\left| c^8_V \right|$ for
$N_{450}$ via Eq.~\eqref{NXfinal} into the expression for
$\ah\,$ of Eq.~\eqref{AFBfinal} yields
\be
\left(1+N_{450}\right)\times A_{FB}^h =\cos2\theta
\left(1.65\,N_{700}-6.15\,N_{450}-19.67 +14.65\sqrt{1.8+N_{450}
-0.24\,N_{700}}\right)\,.
\ee
Plugging in the constraints from Eq.~\eqref{ns}, we find that
the maximal value for $\ah$ is 0.26\,, which is $1.3\,\sigma$
below the mean value in Eq.~\eqref{afb_range}. Regarding the
boosted top cross section, an excess as large as $N_b \sim4$
can be obtained.

\begin{figure}[htb]
\centering
\includegraphics[width=0.45\textwidth]{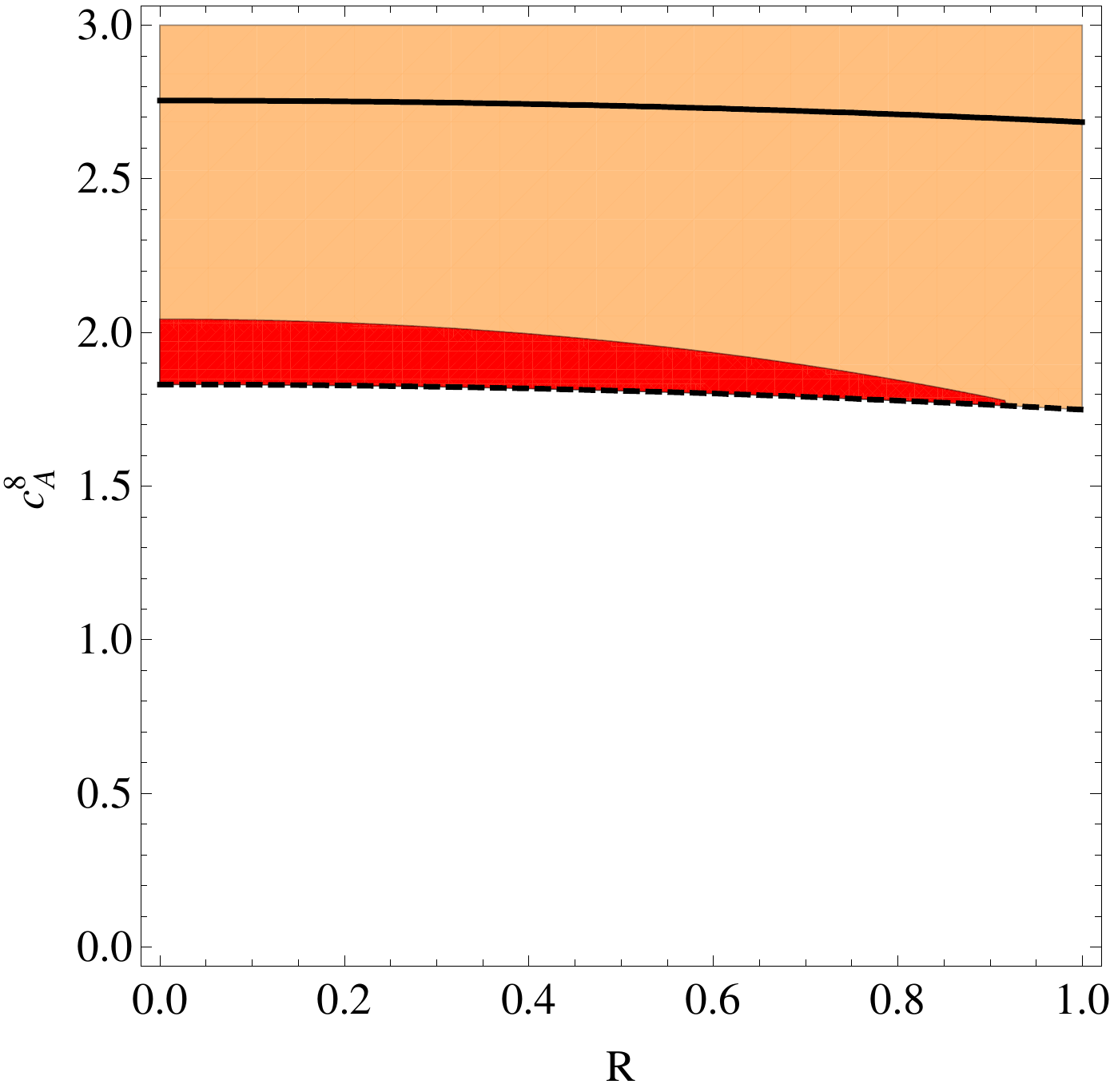}
\caption{The observables under consideration presented in the $R-c_A^8$ plane for $c_V^8=\theta=0\,$:
The solid curve describes the mean value of $\ah$ from Eq.~\eqref{afb_range}, while the
shaded region corresponds to the $1\sigma$ range. The red-shaded region is the overlap of the
latter with the $1\sigma$ constraints on $N_{450}$ and $N_{700}$ in Eq.~\eqref{ns}.}
\label{fig:car}
\end{figure}

\subsection{Axial Interference: $c^8_A \neq 0$ and $c^8_V=0$}
Ref.~\cite{ourfirst} showed that it is possible to explain the
forward-backward asymmetry measurement with only $\cO^8_A\,$.
It is instructive to examine the addition of non-interfering
operators. Fig.~\ref{fig:car} shows the region in the parameter
space of $c^8_A$ and $R$ satisfying Eqs.~\eqref{afb_range}
and~\eqref{ns}. Interestingly, this region is rather narrow,
corresponding to $c_A^8\sim 2$ and
$R \lesssim1\,$. Moreover, only the lower $1\sigma$
range of $\ah$ can be accounted for in this case. (Note that
the central value $\ah \sim0.4$ requires a deviation of
$\sim1.8\sigma$ in $N_{700}\,$, agreeing with~\cite{ourfirst}
modulo the inclusion of $1/\Lambda^{4}$ effects.) As concerns
the high-$p_T$ $t \bar t$ cross section, a maximal excess of
$N_b \sim2$ can be achieved (with $R=0$) along with the
$1\sigma$ range for $\ah\,$.

\subsection{The General Case}
\begin{figure}[tbh]
\centering
\includegraphics[width=0.45\textwidth]{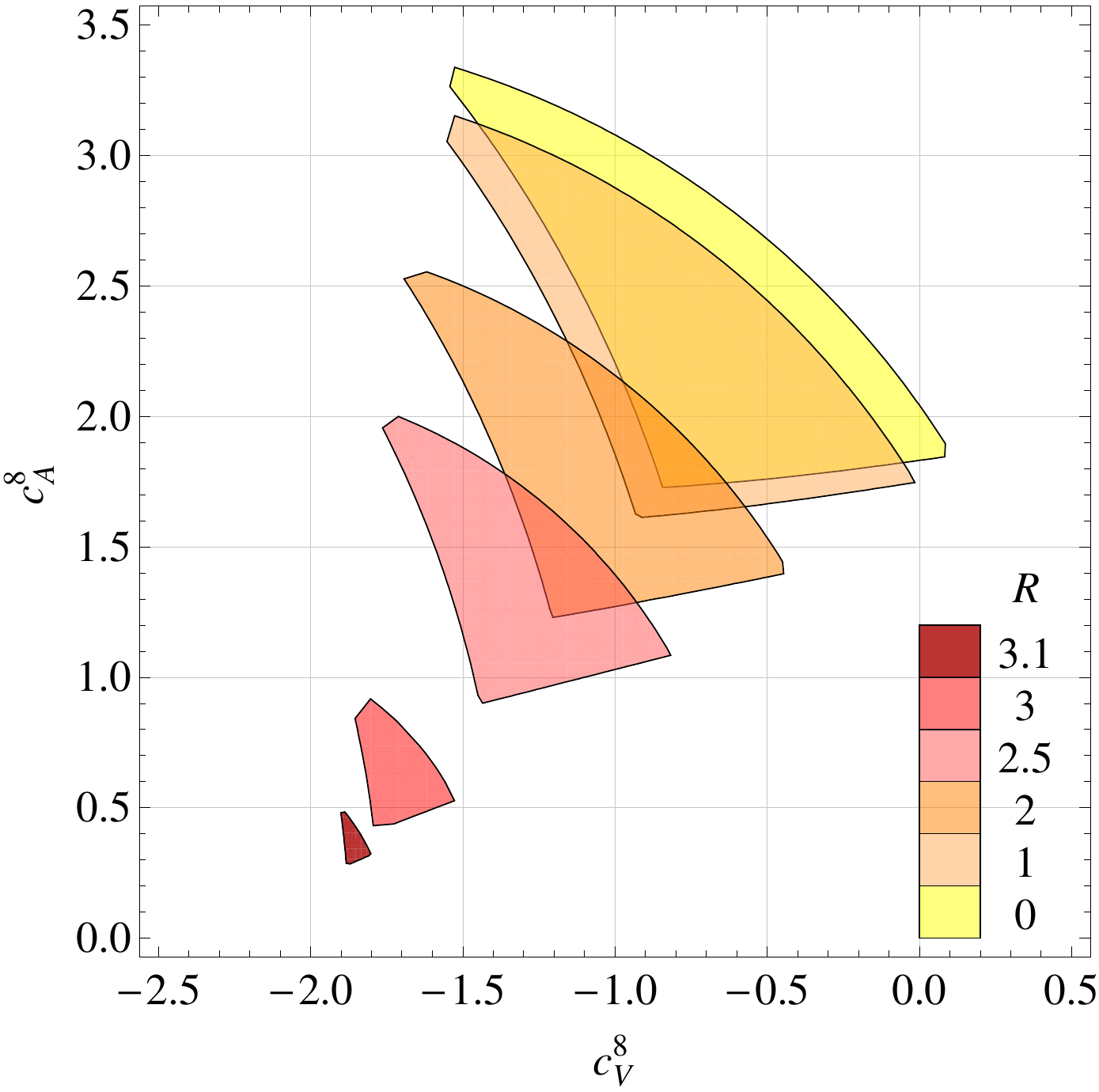}
\caption{The observables under consideration presented in the $c_V^8-c_A^8$ plane: Each region
corresponds to the overlap of the $1\sigma$ ranges for $\ah\,$, $N_{450}$ and $N_{700}$ in
Eqs.~\eqref{afb_range} and~\eqref{ns}, for different values of $R\,$.}
\label{fig:cacv}
\end{figure}

We now explore the general parameter space accounting for the
observables at hand. In Fig.~\ref{fig:cacv} we show the allowed
region in the $c_V^8-c_A^8$ plane for various values of $R\,$.
We learn the following:
\begin{itemize}
\item As $R$ grows, the allowed region becomes smaller, and
    the maximal possible value is $R
    \simeq3.1\,$.
\item The allowed range for the vector octet operator is
    $-2 \lesssim c_V^8 \lesssim0\,$.
\item The allowed range for the axial octet contribution is
    $0.3 \lesssim c_A^8 \lesssim
    3.3\,$.
\end{itemize}
We conclude that:
\begin{itemize}
\item In order to explain the measurement of the $t \bar t$
    forward-backward asymmetry for $\mtt>450$~GeV within
    $1\sigma$, a minimal contribution of the operator $\cO
    _A^8\,$ is necessary, $c_A^8
    \simeq0.3\,$. (This point in the parameter space
    corresponds to $c_V^8 \simeq -1.9$
    and $R \sim 3.1\,$.)
\item The maximal enhancement of the boosted top pair cross
    section is $N_b \sim4\,$. Interestingly, this is
    consistent with $\ah$ within $1\sigma\,$.
\item Accounting for the high mass $t \bar t$
    forward-backward asymmetry within $1\sigma$ dictates a
    minimal excess of $N_b \simeq 0.5\,$.
\item Restricting the parameter space to include only an
    operator of definite chirality for each color
    structure, the maximal $\ah$ is $\sim 0.1\,$. An excess
    of $N_b \sim4$ can still be obtained.
\end{itemize}

\section{Predictions for Near-Future Measurements} \label{sec:pred}
\begin{figure}[hbt]
\centering
\includegraphics[width=0.48\textwidth]{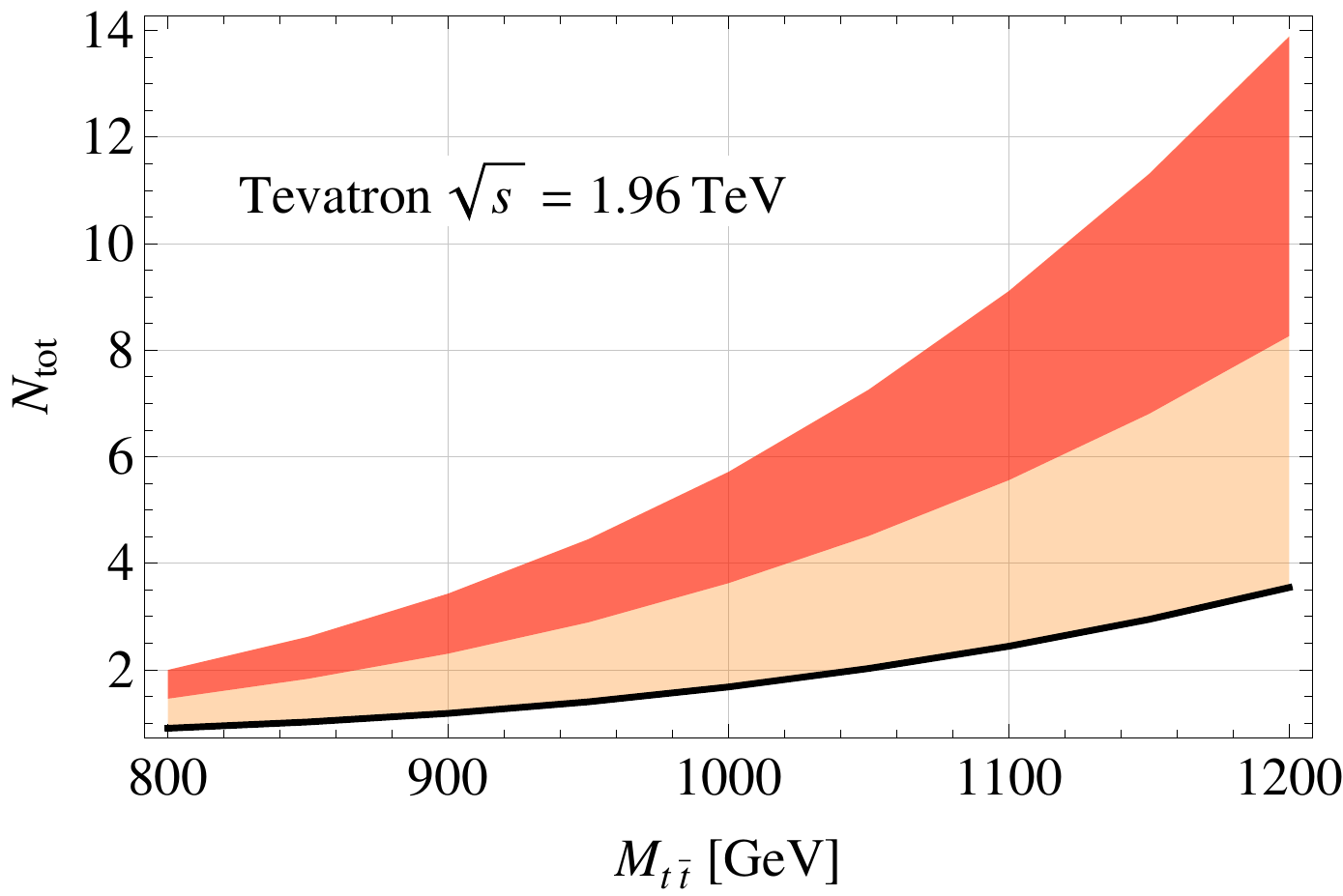} \hfill
\includegraphics[width=0.48\textwidth]{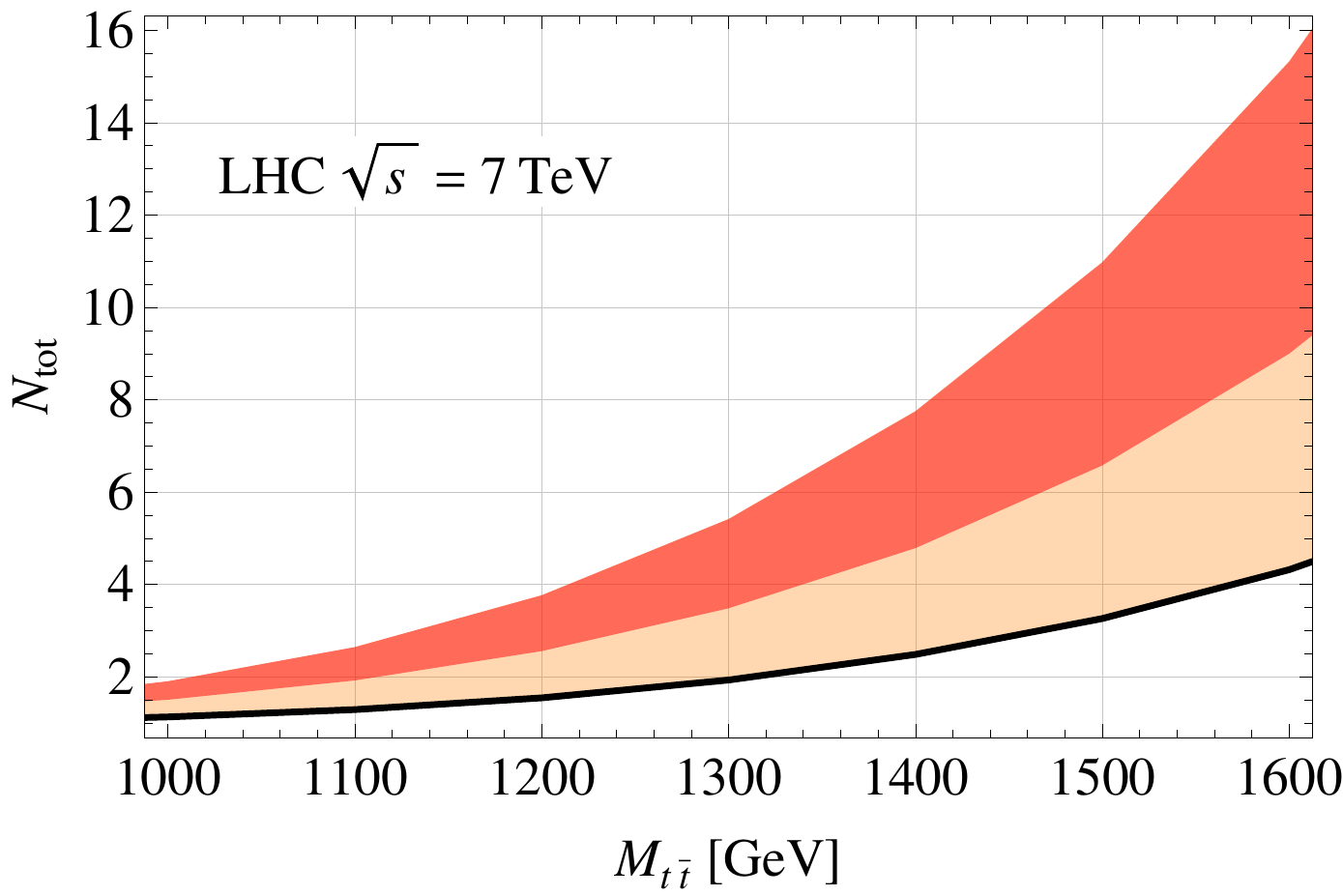}
\caption{The ratio between the total and SM differential cross sections
of top pair production as a function of $\mtt$ at the Tevatron (left) and the
LHC at 7 TeV (right), calculated at leading order. The upper shaded regions
correspond to the mean value of $\ah \sim 0.4\,$, scanning over the allowed range
for $N_{450}$ and $N_{700}$ defined in Eq.~\eqref{ns}. The lower shaded regions correspond
to the lower $1\sigma$ range of $\ah\,$. The thick black curves at the bottom of the shaded
regions correspond to $R=0,\, c_V^8\simeq-0.85,\,c_A^8\simeq1.7\,.$ }
\label{fig:TEVLHC}
\end{figure}

Thus far we focused on existing data from the Tevatron. This
data could have interesting implications on future Tevatron and
LHC measurements. To illustrate this, we consider the $t \bar
t$ differential cross section within our framework. We
emphasize that our results below are general and include in
particular the contributions from scalar and tensor operators.
Fig.~\ref{fig:TEVLHC} depicts the $\mtt$ distribution at the
Tevatron and LHC. The plotted regions correspond to the
predicted enhancement relative to the SM, defined by
\be
N_{\rm tot} \equiv \frac{d\sigma^{\rm SM+NP}/d
\mtt}{d\sigma^{\rm SM}/d \mtt}\, ,
\ee
scanning over the entire parameter space obeying the $1\sigma$
constraints of Eqs.~\eqref{afb_range} and~\eqref{ns}.

We learn that if the latest high mass $t \bar t$
forward-backward asymmetry measurement persists and is
accounted for by heavy new physics, then a significant
enhancement of the $t \bar t$ differential cross section
compared to the SM is expected at both the Tevatron and LHC.
Specifically, at $\mtt \sim 1$~TeV, a minimal factor of two
enhancement is expected at the Tevatron. Similarly, at $\mtt
\sim 1.5$~TeV, the LHC should find a $t \bar t$ cross section
of at least a factor of three higher than within the SM. In
both cases the minimal enhancement is obtained for $R=0$,
namely when only operators interfering with the SM are present.
The combination that minimizes the enhancement, as described by
the thick black curves at the bottom of the shaded regions of
Fig.~\ref{fig:TEVLHC}, does not depend on $\mtt$ and is given
by $c_V^8\simeq-0.85,\,c_A^8\simeq1.7\,.$ To summarize:
\be
\begin{split}
N_{\rm tot}(\mtt=1\,{\rm TeV}) &\gtrsim 2 \qquad \textrm{at the Tevatron}\,,\\
N_{\rm tot}(\mtt=1.5\,{\rm TeV}) &\gtrsim 3 \qquad \textrm{at
the LHC with }\sqrt s=7{\rm \,TeV}\,.
\end{split}
\ee

\section{Outlook}\label{sec:conc}
We have performed a model independent analysis regarding the $t
\bar t$ forward-backward asymmetry, assuming heavy new physics.
Any corresponding high scale new physics model can be mapped to
our formalism to obtain constraints and predictions. We find a
robust prediction in the form of enhancement in hard top
physics at the Tevatron and LHC. The observation of such an
enhancement would be exciting, and our analysis would assist in
interpreting the signal and extracting microscopic information
on the underlying physics. An equally intriguing possibility
would be the absence of such an enhancement, assuming the
asymmetry is established. Our findings would then imply the
presence of sub-TeV new physics. Consequently, the new physics
search strategy should be modified to include precision
analysis in the 100-1000~GeV energy regime.

\section*{Acknowledgments}
We thank Kfir Blum, Alex Kagan, Zohar Komargodski, Seung Lee,
Yosef Nir and Michele Papucci for useful discussions. GP is the
Shlomo and Michla Tomarin career development chair and is
supported by the Israel Science Foundation (grant \#1087/09),
EU-FP7 Marie Curie, IRG fellowship, Minerva and G.I.F., the
German-Israeli Foundations, and the Peter \& Patricia Gruber
Award.

\end{document}